\documentclass[a4paper,11pt]{article}
\usepackage{pos}

\usepackage{hyperref}
\usepackage{booktabs}
\usepackage{lineno}
\usepackage{color}
\usepackage{tabularray}
\usepackage{cleveref} 

\title{Multi-parton interactions in pp collisions using charged-particle flattenicity with ALICE}

\author*[a]{Gyula Benc\'edi}
\onbehalf{on behalf of the ALICE Collaboration}

\affiliation[a]{HUN-REN Wigner Research Centre for Physics,\\
  29-33 Konkoly-Thege Miklos, Budapest, Hungary}

\emailAdd{bencedi.gyula@wigner.hun-ren.hu}

\abstract{
Event classifiers based either on the charged-particle multiplicity or on event topologies, such as spherocity and underlying event, became very useful tools to study collective-like behaviors in small collision systems. However, multiplicity-based event classifiers were shown to bias the data sample in a way that can obscure the effects of multi-parton interactions, and, this way, make it difficult to pin down the origins of small-system collectivity.\\

In this proceedings, the measurement of the transverse momentum ($p_{\mathrm{T}}$) spectra of primary charged pions, kaons, (anti)protons and unidentified hadrons in inelastic pp collisions at $\sqrt{s}=13~\mathrm{TeV}$ are reported. Events are classified using a novel event shape observable, flattenicity, that was proposed to select minijet-enhanced pp collisions. Particle production is studied as a function of flattenicity and double-differentially as a function of flattenicity and charged-particle multiplicity. The results are compared with theoretical predictions from the PYTHIA~8 and EPOS~LHC Monte Carlo models.
}

\FullConference{The European Physical Society Conference on High Energy Physics (EPS-HEP2023)\\
 21-25 August 2023\\
Hamburg, Germany\\}

\begin{document}
\maketitle

\section{Introduction}


Despite the large amount of soft-QCD results on collectivity in pp and p--Pb collisions (small-collision systems), the origin of these phenomena is not yet fully understood; however, in this direction, several attempts were made by the ALICE Collaboration, see e.g. Ref.~\cite{ALICE:2022qxg}. It is noteworthy that amongst various theoretical approaches, the PYHTIA~8~\cite{Skands:2014pea} model can reproduce many of the experimental results. It does so by the modeling of a non-perturbative final-state effect, color reconnection (CR), and multi-parton interactions (MPI). MPI allow the simultaneous occurrence of several incoherent binary semi-hard partonic interactions in a single pp collision, which produce multi-minijets topologies. When the event classification is performed in charged-particle multiplicity measured at large pseudorapidities~\cite{ALICE:2020nkc}, the multiplicity based on PYTHIA is strongly correlated with the MPI activity~\cite{Ortiz:2022mfv}.

In order to increase the sensitivity to MPIs in the measurements, an event classifier $R_{\rm T}$ is constructed~\cite{Martin:2016igp} to measure the charged-particle multiplicity normalised to its event-averaged value in the underlying event (UE) region based on the CDF method~\cite{Charged_Jet_evolution_and_UE_pbarp_1.8TeV}. Measurements from ALICE indicate that the spectral shapes of charged particles experience a hardening with increased MPI activity in the UE-dominated topological region~\cite{ALICE:2023csm}. This effect can be explained by a selection bias towards multijet topologies due to soft gluon radiation~\cite{RT_TransMax_TransMin_Antonio_Gyula}.

A recent study has explored an event shape observable with a strong sensitivity to soft MPIs and CR effects using a multivariate regression technique~\cite{Ortiz:2020rwg}. The ratio of the yield of charged pions in MPI-enhanced pp collisions to that in minimum-bias (MB) pp collisions showed a pronounced peak in the $p_{\mathrm{T}}$ region of $1<p_{\rm T}<8\,\mathrm{GeV}/c$, which has not been observed in pp data before.

In this contribution, a novel event classifier, flattenicity~\cite{Ortiz:2022mfv}, that quantifies the shape of the event using experimental information from both azimuthal and forward/backward pseudorapidity directions is explored. The $p_{\mathrm{T}}$ spectra are studied as a function of flattenicity and double-differentially as a function of flattenicity and charged-particle multiplicity in pp collisions at $\sqrt{s}=13$\,TeV.

\section{Event classification using charged-particle flattenicity}

A MB data sample of about $1.64\times10^{9}$ from pp collisions at a centre-of-mass energy of $\sqrt{s}=13$\,TeV is used to measure the production of primary charged particles ($\mathrm{h}^{\pm}$) and charged pions ($\pi^{\pm}$), kaons ($\mathrm{K}^{\pm}$), and (anti)protons ($(\overline{\mathrm{p}})\mathrm{p}$). The MB trigger requires a charged-particle signal in the V0 detectors, covering the pseudorapidity regions $2.8<\eta<5.1$ (V0A) and $-3.7<\eta<-1.7$ (V0C). Both the V0A and V0C detectors contain four rings in the $\eta$ direction and eight equidistant sectors in the azimuthal direction, resulting in a grid of $N_{\mathrm{cell}}=64$ cells in their acceptance~\cite{ALICE:2013axi}. The silicon-based inner tracking system and the time projection chamber (TPC) are used for tracking, while the TPC and the time-of-flight detector provide particle identification in $|\eta|<0.8$ and in the $p_{\mathrm{T}}$ range of $0.15<p_{\mathrm{T}}<20\,\mathrm{GeV}/c$, depending on partilce species.

The accepted events are required to have at least one charged particle produced in $|\eta|<1$. The $p_{\mathrm{T}}$ spectra of charged and identified particles are measured as a function of the charged-particle multiplicity and flattenicity~\cite{Ortiz:2022mfv}. Particle multiplicites are measured by signal amplitudes in the V0A and V0C detectors; this classification is denoted as V0M. Flattenicity is quantified on an event-by-event basis as follows: $\rho=\sqrt{\sum_{i}{(N_{\rm ch}^{{\rm cell},i} - \langle N_{\rm ch}^{\rm cell} \rangle)^{2}}/N{_{\mathrm{cell}}^{2}}}/\langle N_{\rm ch}^{\rm cell} \rangle$, where $N_{\rm ch}^{{\rm cell},i}$ is the average multiplicity in the $i$-th cell, $\langle N_{\rm ch}^{\rm cell} \rangle$ is the average of $N_{\rm ch}^{{\rm cell},i}$ in the event. To easier associate flattenicity with more inclusive event shape observables, the results are presented as a function of $1-\rho$: multi-minijet topologies yield small flattenicity values ($1-\rho\rightarrow 1$), whereas events dominated by multijet topologies have large flattenicity values ($1-\rho\rightarrow 0$). The average charged-particle pseudorapidity densities $\langle{\rm d}N_{\rm ch}/{\rm d}\eta\rangle$ measured in $|\eta| < 0.8$ for the different flattenicity classes show a clear correlation with $1-\rho$: multijet events (50-100\% $1-\rho$) have lower  $\langle {\rm d}N_{\rm ch}/{\rm d}\eta \rangle$ than multi-minijet (0-1\% $1-\rho$) ones. The implicit multiplicity dependence can be factorized by performing a double-differential analysis, i.e. the flattenicity selection is also performed for high-multiplicity ($0-1$\% V0M class) events. The particle identification is performed using the standard techniques applied in previous ALICE measurements, see recently in Refs.~\cite{ALICE:2020nkc,ALICE:2023yuk}.

\section{Results and Discussion}

\begin{figure}[t]
    \centering
    \hspace{0cm}
    \includegraphics[width=1\textwidth]{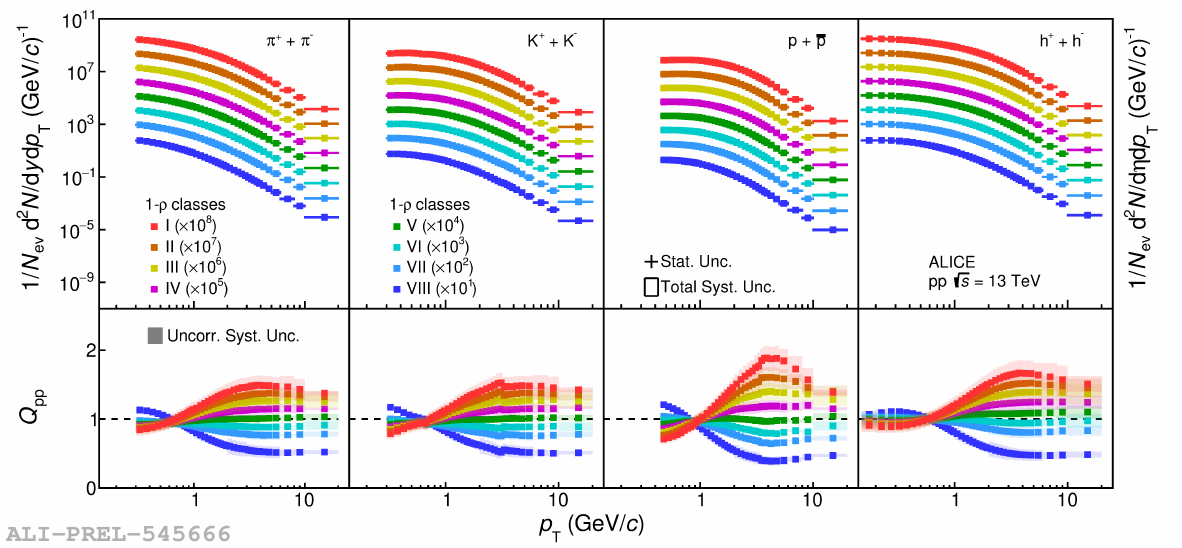}\\
    \includegraphics[width=1\textwidth]{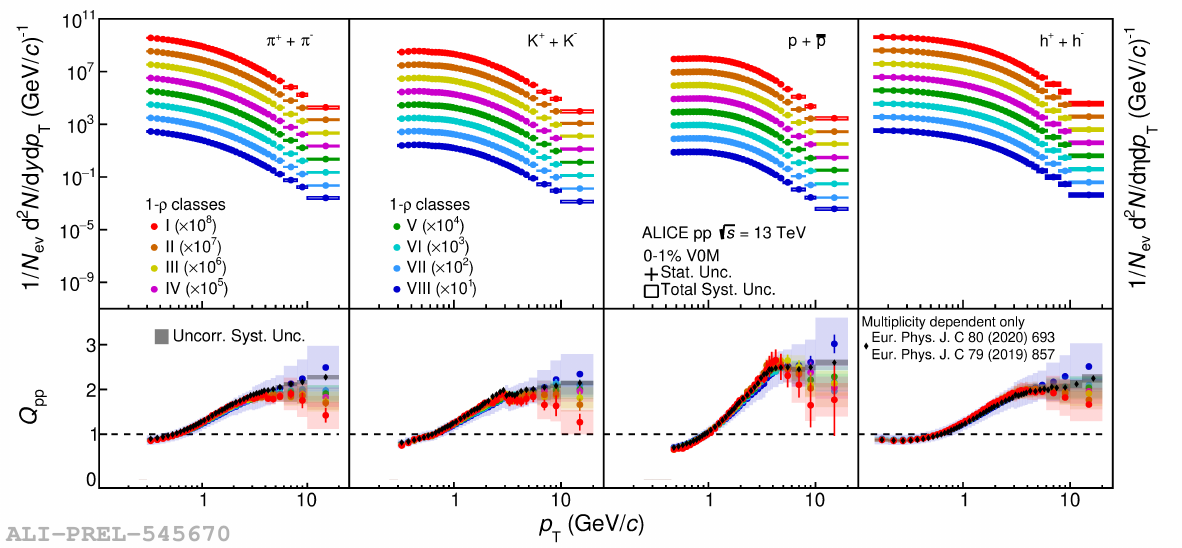}
    \hspace{0cm}
    \caption{Transverse momentum ($p_{\rm T}$) spectra of $\pi^{\pm}$, $\mathrm{K}^{\pm}$, $(\overline{\mathrm{p}})\mathrm{p}$, and $\rm{h}^{\pm}$ for different flattenicity event classes (top figure), and for high-multiplicity events (0-1\% V0M) in the same flattenicity event classes (bottom figure). Bottom panels in each figure show the $Q_{\rm pp}$ for the corresponding event classes. The multiplicity dependent $Q_{\rm pp}$ values are taken from Refs.~\cite{ALICE:2019dfi,ALICE:2020nkc}.}
    \label{fig:spectra_with_rpp}
\end{figure}

The analyses of the $p_{\mathrm{T}}$ spectra are performed using standard methods~\cite{ALICE:2019dfi,ALICE:2020nkc,ALICE:2023yuk}. Figure~\ref{fig:spectra_with_rpp} shows $p_{\rm T}$ spectra of $\pi^{\pm}$, $\mathrm{K}^{\pm}$, $(\overline{\mathrm{p}})\mathrm{p}$, and $\rm{h}^{\pm}$ for different $1-\rho$ flattenicity event classes (top figure), and for events in the 0-1\% V0M multiplicity and flattenicity classes (bottom figure). The evolution of the $p_{\mathrm{T}}$-spectral shapes with flattenicity can be quantified by the ratio ($Q_{\mathrm{pp}}$) of particle yield measured in a given $1-\rho$ class normalized to the yield measured in MB pp collisions: $Q_{\mathrm{pp}} = (\mathrm{d}^{2}N/\langle \mathrm{d}N_{\mathrm{ch}}/\mathrm{d}\eta \rangle \mathrm{d}y\mathrm{d}p_{\mathrm{T}})^{1-\rho}/(\mathrm{d}^{2}N/\langle \mathrm{d}N_{\mathrm{ch}}/\mathrm{d}\eta \rangle \mathrm{d}y\mathrm{d}p_{\mathrm{T}})^{\mathrm{minimum~bias}}$. The ratio is scaled by $\langle \mathrm{d}N_{\mathrm{ch}}/\mathrm{d}\eta \rangle^{1-\rho}/\langle \mathrm{d}N_{\mathrm{ch}}/\mathrm{d}\eta \rangle^{\mathrm{minimum\,bias}}$ that is sensitive to the average number of MPIs, according to PYTHIA. Going from multijet (50-100\% $1-\rho$, class VIII) to multi-minijet (0-1\% $1-\rho$, class I) topologies, events on average have from about half to three times larger $\langle{\rm d}N_{\rm ch}/{\rm d}\eta\rangle$ with respect to MB events. A clear development of a peak structure for the event class I is observed for $1<p_{\mathrm{T}}<8~\mathrm{GeV}/c$, and the maximum of the peak shows a mass-dependent ordering that can be attributed to radial flow~\cite{ALICE:2020nkc,ALICE:2023yuk}. The bottom part of Fig.~\ref{fig:spectra_with_rpp} reports the results from a double differential analysis: the high-multiplicity 0-1\% V0M event class (having on average 3-4 times larger $\langle{\rm d}N_{\rm ch}/{\rm d}\eta\rangle$ with respect to MB events) reveals similarity to the flattenicity-only case. However, the $Q_{\mathrm{pp}}$ in the event class VIII increases over the entire $p_{\mathrm{T}}$ range. The effect was also seen for V0M-only event selections~\cite{ALICE:2023yuk,ALICE:2020nkc}, and it is a consequence of jet bias~\cite{ALICE:2019dfi}.

\begin{figure}[t!]
    \centering
    \hspace{0cm}
    \includegraphics[width=1\textwidth]{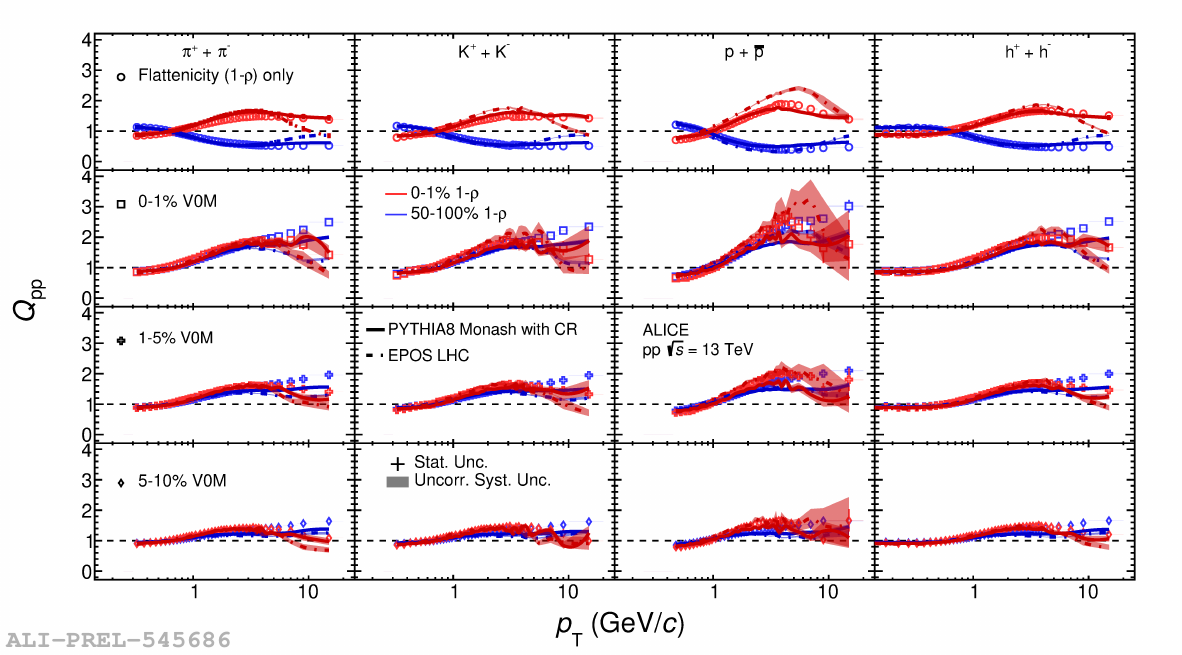}
    \hspace{0cm}
    \caption{The $Q_{\rm pp}$ of $\pi^{\pm}$, $\mathrm{K}^{\pm}$, $(\overline{\mathrm{p}})\mathrm{p}$, and $\rm{h}^{\pm}$ for the 0-1\% and 50-100\% $1-\rho$ classes. The data are compared with model predictions from PYTHIA~8 and EPOS~LHC. The statistical and uncorrelated systematic uncertainties are represented with bars and shaded areas, respectively.}
    \label{fig:rpp_with_models}
\end{figure}

Figure \ref{fig:rpp_with_models} compares the measured $Q_{\mathrm{pp}}$ with model predictions from EPOS-LHC~\cite{EPOS_LHC} and PYTHIA~\cite{Skands:2014pea} (with CR) including detector effects, where only the extreme flattenicity selections are examined, 0-1\% and 50-100\% $1-\rho$, inclusively and for high-multiplicity events. PYTHIA 8.3 with the Monash 2013 tune including the MPI and CR models with default parameter sets generally describes the data presented in flattenicity event classes. In contrast, EPOS LHC with parametrized collective hydrodynamics describes the data only partially (low-to-mid $p_{\rm T}$), while at high $p_{\rm T}$ an opposite trend is seen with respect to PYTHIA.

\begin{figure}[t!]
    \centering
    \hspace{0cm}
    \includegraphics[width=1\textwidth]{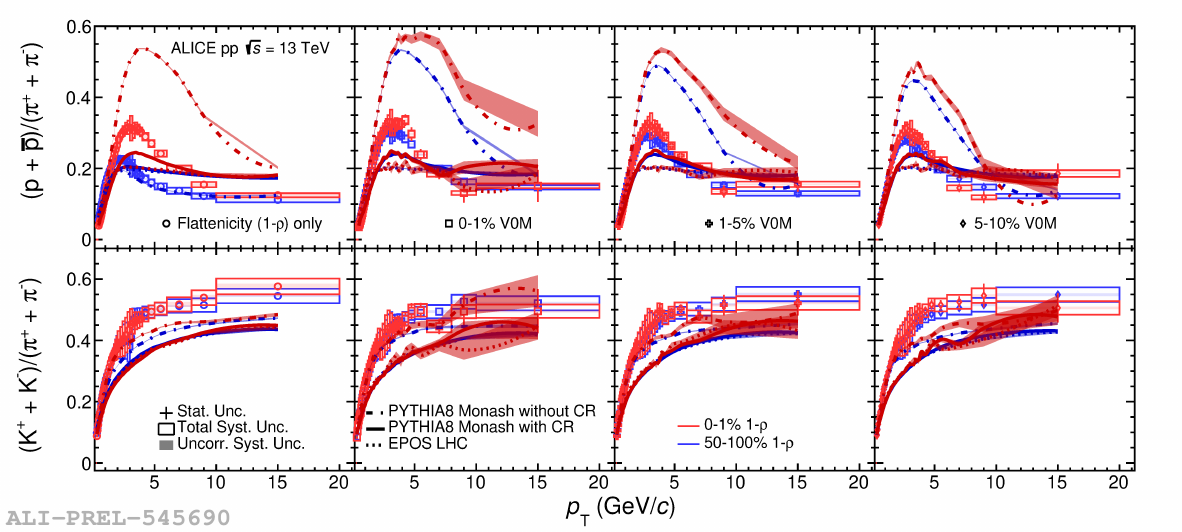}
    \hspace{0cm}
    \caption{Proton-to-pion (top row) and kaon-to-pion (bottom row) particle ratios as a function of $p_{\rm T}$ for the 0-1\% and 50-100\% $1-\rho$ classes, and for various V0M event classes. The data are compared with model predictions from PYTHIA~8 and EPOS~LHC, where the shaded bands represent statistical uncertainties.}
    \label{fig:particle_ratios_with_models}
\end{figure}

Figure~\ref{fig:particle_ratios_with_models} shows the $p_{\rm T}$-differential proton-to-pion ($\mathrm{p}/\pi$) and kaon-to-pion ($\mathrm{K}/\pi$) ratios for the two extremes of flattenicity, 0-1\% and 50-100\% $1-\rho$, inclusively and for high-multiplicity events. The $\mathrm{K}/\pi$ ratio does not change neither with flattenicity, nor with V0M multiplicity selections. The models follow this trend qualitatively. But the $\mathrm{p}/\pi$ ratio is sensitive to flattenicity-only selection. The data is described by PYTHIA (with CR): the model predicts the enhanced baryon-to-meson ratio witnessed for the 0-1\% $1-\rho$ event class. For the double-differential analysis, the data indicates similar $\mathrm{p}/\pi$ ratios between the two extremes of flattenicity classes. The measurement is supported by PYTHIA, whereas EPOS LHC fails to describe the observed trends. Notably, the worst description of the data by the models is provided in the highest multiplicity (0-1\% V0M) event class.

\section{Conclusion}

ALICE studied a novel event shape observable flattenicity in pp collisions at $\sqrt{s}=13~\mathrm{TeV}$ for the first time. For multi-minijet events, the ratio of event-class dependent $p_{\rm T}$ spectra to that of MB ($Q_{\mathrm{pp}}$) develops a pronounced peak with increasing multiplicity that is mass dependent. Results are qualitatively described by the PYTHIA model based on color strings and indicate that flattenicity-selected events show reduced sensitivity to multijets.

\acknowledgments
Supported by Hungarian National Research, Development and Innovation Office (NKFIH) grants OTKA PD143226, FK131979, K135515, 2021-4.1.2-NEMZ\_KI-2022-00018/11223-23, and by CONAHCyT under the grant A1-S-22917.

\bibliographystyle{JHEP}
\bibliography{bibliography}

%
%
\end{document}